\documentclass[prb,twocolumn,showpacs]{revtex4}

\usepackage{graphicx}
\usepackage{dcolumn}
\usepackage{amsmath}

\begin{document}

\title{From the Square Lattice to the Checkerboard Lattice : \\
Spin Wave and Large-n limit Analysis}



\author{Benjamin Canals}
\email{canals@polycnrs-gre.fr}
\homepage{http://ln-w3.polycnrs-gre.fr/pageperso/canals/}
\affiliation{Laboratoire Louis N\'eel, CNRS, 25 avenue
des Martyrs, Boite Postale 166, 38042 Grenoble Cedex 9, France}

\date{\today}

\begin{abstract}
Within a spin wave analysis and a fermionic large-n limit, it is shown that the antiferromagnetic 
Heisenberg model on the checkerboard lattice may have different ground
states, depending on the spin size $S$.
Through an additional exchange interaction that corresponds to an 
inter-tetrahedra coupling, the stability of the N\'eel state has been
explored for all cases from the square 
lattice to the isotropic checkerboard lattice.
Away from the isotropic limit and within the linear spin wave
approximation, it is shown that there exists 
a critical coupling for which the local magnetization of the N\'eel
state vanishes for any value of the spin $S$.
One the other hand, using the Dyson-Maleev approximation, this result 
is valid only in the case $S= \frac12$ and the limit between a 
stable and an unstable N\'eel state is at $S=1$.
For $S= \frac12$, the fermionic large-n limit suggests that the ground
state is a valence bond solid build with disconnected 4-spins singlets.
This analysis indicates that for low spin and in the isotropic limit, 
the checkerboard antiferromagnet may be close to an instability between
an ordered $S=0$ ground state and a magnetized ground state.
\end{abstract}

\maketitle

\section{Introduction\label{introduction}}
In one dimension (1D) the ground state of the Heisenberg antiferromagnet may be 
quasi-ordered or disordered, depending on the spin parity S \cite{haldane}.
When S is an integer, and in particular, when $S=1$, the ground state 
corresponds to a magnetically disordered state where spin-spin correlation
functions (CFs) exponentially decay with distance and with total spin 
$S_{\rm{t}}=0$.
This state is called a quantum spin liquid (QSL) as the spin-spin
CFs are equivalent to the density-density CFs in conventional 
liquids and because it has total spin $S=0$ and no broken symmetry.
This concept can be generalized to classical systems if the
spins are considered as $O(3)$ vectors.
The constraint $S=0$ is then removed and the state is termed
a classical spin liquid (CSL).
Another notion for spin liquid ground states emerged from earlier
studies by Wannier \cite{wannier50} and Houtappel \cite{houtappel50}
and is related to geometrical frustration.
In that case, it is the geometry of the underlying lattice that leads
to magnetic disorder and not its dimensionality.
Naturally, frustrated structures may exist not only for $D=1$ and
for over 50 years frustrated lattices have been examined in the
search for disordered ground states in higher dimensions ($D=2,3$).
One goal of this research into peculiar structures is the study of
unconventionnal spectra, not covered by the 
Lieb and Mattis theorem \cite{liebmattis62}.

\begin{figure}
\includegraphics[width=7.5cm]{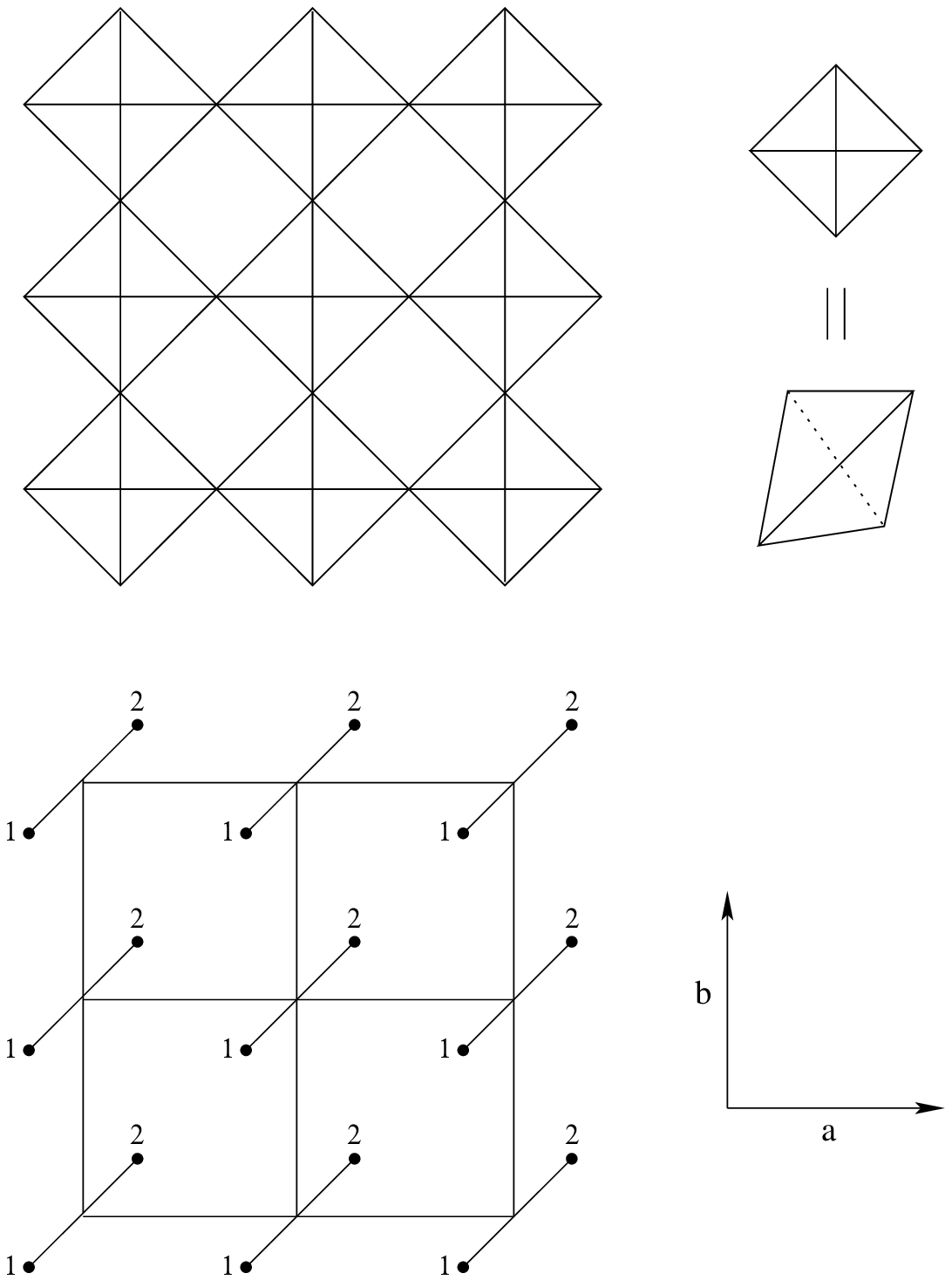}%
\caption{The checkerboard lattice. It can be pictured as a square lattice
of tetraheda (top). Locally, it reproduces the same environment
as the three dimensional pyrochlore lattice. In order to simplify our
calculations, it has been described by a square 
lattice with a 2-spins unit cell (bottom).
\label{checkerboard_lattice}}
\end{figure}

To date, two particular lattices have been considered
as good candidates for spin liquid ground states : 
the kagom\'e lattice and the pyrochlore lattice.
The first one is a two dimensional arrangement of corner sharing triangles,
while the latter is a three dimensional structure of corner sharing
tetrahedra.
Both lattices have been theoretically investigated and it has been suggested
that in the classical case, as well as in the quantum case, they possess
spin liquid like ground states \cite{leche97,garcan99,moecha98,canals9800}.
Experimentally, all systems that can be well described by the antiferromagnetic
Heisenberg model on these lattices have been found to possess this spin 
liquid behavior or unconvential behaviors that can be ascribed in part to 
the underlying ``theoretical'' spin liquid ground state 
\cite{diep94}.
This suggests that both the elementary cells (triangle or tetrahedron) and 
the connectivity (corner sharing) are major ingredients for
magnetic disorder.
In this paper, we study such a lattice, where the elementary cell is a 
tetrahedron and where connectivity is corner sharing~:~the checkerboard
lattice (see Fig.~\ref{checkerboard_lattice}).
This system is two dimensionnal and can be described as the two dimensional
analog of the pyrochlore lattice : each atom possess locally the same
environment.
It has recently been proven that its quantum ground states are 
singlets \cite{lieb99} for finite systems.
The aim of this work is to test wether this conclusion may be valid for
an infinite lattice.
To do so we start away from the isotropic checkerboard limit, 
when diagonal bonds are weakened (see Fig.~\ref{check-and-env}),
and we test wether this system undergoes a Neel ordering transition
when reaching the isotropic limit.
This approach is very close in spirit to the one previously done by
Chandra and Dou\c cot \cite{chandra88} on the $J_1$-$J_2$ square lattice
model.
This is done within the frame of the linear spin wave theory and the
Dyson-Maleev approximation \cite{dyson56}.
Whereas for the linear spin wave approach there always exists a critical
coupling for which the local magnetization is unstable, whatever the value
of the spin, in the case of the Dyson-Maleev approach there is a critical
coupling only for the case $S=\frac12$.
The peculiar case $S=\frac12$ is explored within a fermionic large-n limit 
of the SU(2) hamiltonian.
It is found that the system should order in a valence bond solid of 
disconnected 4-spins singlets.
The notations, the hamiltonian, and the linear spin wave analysis 
are introduced in Sec. \ref{hamiltonian_and_classical_analysis}.
The Dyson-Maleev formalism is shortly described in Sec. \ref{dysonmaleev}
as it is can be seen as an extension of Sec. \ref{hamiltonian_and_classical_analysis}.
Results of the linear spin wave theory and the Dyson-Maleev approach are 
presented and discussed in Sec. \ref{results-sw}.
The case $S=\frac12$ is detailed in Sec. \ref{s-small}.

\section{Lattice, Hamiltonian and Spin Wave Analysis
\label{hamiltonian_and_classical_analysis}}
\subsection{Notations and Lattice Description
\label{notations_and_lattice_description}}
\begin{figure}
\includegraphics[width=8.5cm]{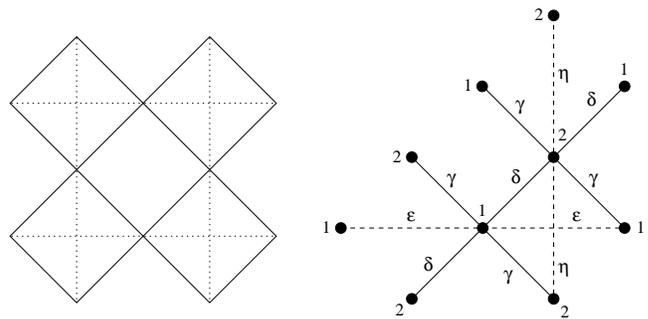}%
\caption{(left) Lattice of coupling constants $J$ and $J'$. Bold lines
design constant $J$ while dotted lines describe constant $J'$.
(right) Local environment of a unit cell. $\vec{\delta}$ and $\vec{\gamma}$
are related to $J$-like coupling while $\vec{\epsilon}$ and $\vec{\eta}$ 
correspond to $J'$.
\label{check-and-env}}
\end{figure}

The checkerboard lattice can be described as a square lattice with
a unit cell containing 2 spins (see Fig.~\ref{checkerboard_lattice}).
Each cell of the lattice can be obtained from the others 
by applying the translations $a=(1,0)$ and $b=(0,1)$.
This means that each site of the lattice can be defined
by two indices $(i,n)$.
The first corresponds to the unit cell ($i=1 \dots \rm{N}$)
where N is the total number of cells in the lattice
and the other to the type of the site ($n=1,2$).  
We introduce two kinds of interactions, $J$ and $J'$. 
$J$ corresponds to a coupling constant on a square lattice. 
$J'$ can be seen as a coupling constant within one square over two of the
square lattice 
(see Fig.~\ref{check-and-env}).
Using these notations, the Brillouin zone of the underlying
Bravais lattice is 
$\rm{BZ}(k_{x} , k_{y})=[- \pi , \pi] \times [- \pi , \pi]$ 
and the structure coefficients can be defined as 
$\delta = \cos(\vec{k}.\vec{\delta}) = \cos \left( (k_{\rm{x}} + k_{\rm{y}})/2 \right)$,
$\gamma = \cos( \vec{k}.\vec{\gamma}) = \cos \left( (k_{\rm{x}} + k_{\rm{y}})/2 \right)$,
$\epsilon = \cos(\vec{k}.\vec{\epsilon}) = \cos(k_{\rm{x}})$,
$\eta = \cos(\vec{k}.\vec{\eta}) = \cos(k_{\rm{y}})$
where $\vec{\delta}$, $\vec{\gamma}$, $\vec{\epsilon}$ and $\vec{\eta}$ are 
depicted in Fig.~\ref{check-and-env}.
\subsection{Spin Wave Hamiltonian
\label{sw-hamiltonian}}
\begin{figure}
\hspace{-1cm}
\vspace{-0.5cm}
\includegraphics[width=7cm]{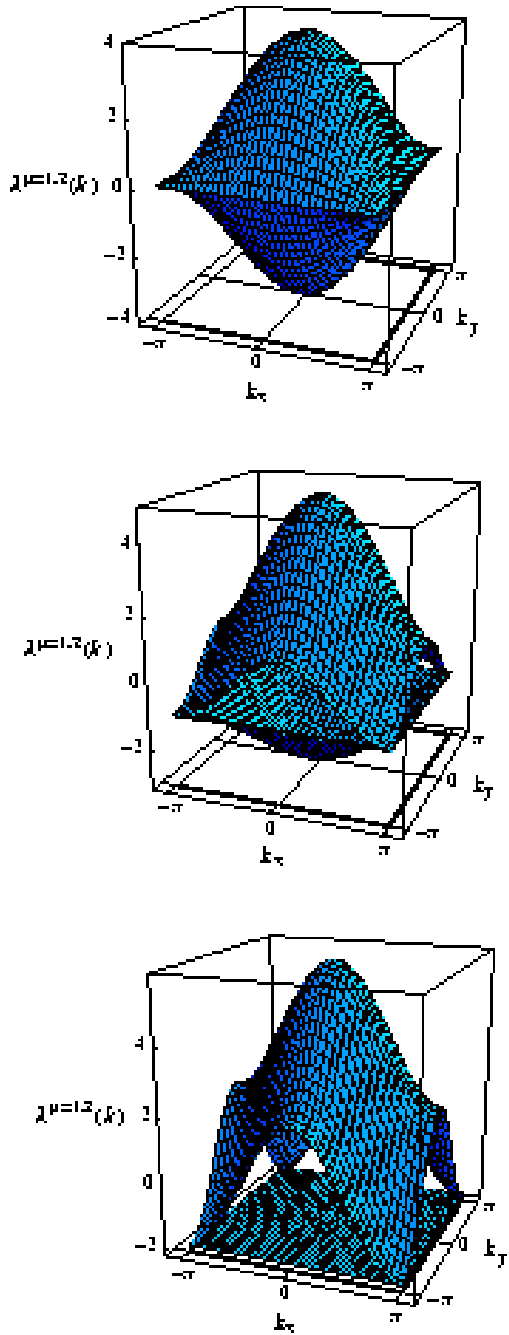}%
\caption{Fourier transform of the interactions for the square
lattice limit ($r=0$, top), an intermediate case ($r=0.5$, middle)
and the checkerboard limit ($r=1$, bottom).
Each manifold corresponds to an eigenvalue of $J(\vec{q})$.
\vspace{-0.5cm}
\label{fourier-interactions}}
\end{figure}

\begin{figure}
\hspace{-1cm}
\includegraphics[width=7cm]{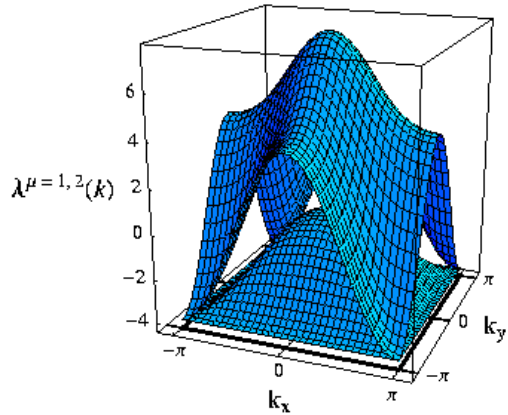}%
\vspace{+0.1cm}
\caption{Fourier transform of the interactions for $r=2$.
Each manifold corresponds to an eigenvalue of $J(\vec{q})$.
The mean field ground state is continuously degenerate along
lines which unable the choice of a trial ground state for the
spin wave analysis.
\label{rgreater1}}
\end{figure}

The Heisenberg hamiltonian for this system is 
\begin{equation}
\label{hamiltonian}
H = - \sum_{ij} J_{ij} S_i . S_j
\end{equation}
\noindent
where $J_{ij}$ are coupling constants ($J_{ij}<0$ for an antiferromagnet) and 
$S_i$ is the spin located at site $i$.
This spin hamiltonian may be transformed into a boson hamiltonian through an
Holstein-Primakoff transformation, provided that we
know its Mean Field ground states (MF) from the square lattice limit to the 
checkerboard lattice limit, i.e for all values of $r=J'/J$.
One way to know the MF ground state(s) is to compute the Fourier transform 
of the interactions on the lattice \cite{rei91}.
\begin{equation}
H = - \sum_{i,j} J_{ij} S_i . S_j
  =   \sum_{\vec{q}} J(\vec{q}) S_{\vec{q}}.S_{-\vec{q}}
\end{equation}
As we here deal with a 2-spin unit cell, its Fourier transform 
$J(\vec{q})$ is a  $2 \times 2$ matrix and must be diagonalized
to find the normal modes $\lambda^{\mu =1,2} (\vec{q})$.
The ground state (if unique) then corresponds to the lowest eigenvalue 
$\lambda^{1} (\vec{q_0})$ and is described
by the propagation vector $\vec{q}_0$ and the associated 
eigenvector $V_{1} (\vec{q}_0)$
which fixes the inner structure of the unit cell.
In the range $0 \leq r < 1$, $\vec{q}_0$ always equals $0$ and
the eigenvector is always equal to $(1,-1)$ which means that the 
ground state is non-degenerate and corresponds to the 
antiferromagnetic N\'eel state (see Fig.~\ref{fourier-interactions} 
top and middle).
At the peculiar point $r=1$, we obtain a flat branch, signaling that
at the MF level, the system is continuously degenerate 
(see Fig.~\ref{fourier-interactions} bottom), as previously noted in 
Ref. \onlinecite{rei91}.
That behavior can be explained in terms of local degrees of freedom.
In the limit $J'=J$, we note
that the hamiltonian may be decomposed as a sum of squares over 
``tetrahedra'', as it has been done in the kagom\'e 
as well as in the pyrochlore lattice
\begin{eqnarray}
\label{constraints}
H & = & - J \sum_{ij} S_i . S_j \nonumber \\ 
  & = & - \frac{J}{2} \sum_{\rm{Tet.}} \left( \sum_{i=1}^{4} S_i  \right)^2 
        + 2 \rm{N} J S (S+1)
\end{eqnarray}
We see that the classical ground state is defined by the constraint 
$\sum_{i=1}^{4} S_i = 0$ on each tetrahedron.
Furthermore, as tetrahedra are corner sharing, this constraint is not
sufficient to order the system and there are an infinite number of
ways to satisfy it, which results in a flat branch over the whole
Brillouin zone.
This is exactly the same phenomenon that was previously observed on
the kagom\'e as well as on the pyrochlore lattice.
Nevertheless, the N\'eel state belongs to this infinite set of
MF ground states, and as it is the correct one for 
any $r<1$, we will also consider this state by continuity for the point $r=1$.
For the parameter range $r>1$, the situation is quite different as MF ground states
are continuously degenerate along lines corresponding to the border of the square 
lattice Brillouin zone (see Fig.~\ref{rgreater1}).
This precludes a description via a spin wave analysis in this range as no magnetic
ground state can be choosen as a trial state.
This is the same situation as the one encountered on the kagom\'e or the pyrochlore
lattices where spin wave calculations can only compare relative stability of 
N\'eel states without considering the entire set of these states \cite{chubukov92}.
So, we will restrict ourselves to the range $0 \leq r \leq 1$, and consider 
the antiferromagnetic N\'eel state as the trial state.
We therefore introduce two species of bosons, each one related to a different sublattice,
and restrict the developpement of spin operators to the lowest order in the spin
wave expansion
The resulting bosonic hamiltonian is then diagonalized through a Bogoliubov transformation.
We are left with two branches of ``magnons'' whose dispertions are
$\epsilon_\alpha = (1/2) (f - g + \sqrt{(f + g)^2 - 4 h^2})$
and
$\epsilon_\beta = (1/2) (-(f - g) + \sqrt{(f + g)^2 - 4 h^2})$
where $f_k = 2+r(\epsilon - 1)$, $g_k = 2+r(\eta - 1)$, $h_k = \delta + \gamma$ and
$r=J'/J$ ($\delta$, $\gamma$, $\epsilon$ and $\eta$ are defined 
in Sec.~\ref{notations_and_lattice_description}).

Restricting to the ground state, the energy as well as the local magnetization are
obtained through
\begin{eqnarray}
\label{energy-discrete}
E & = & \frac{1}{2\rm{N}} H = 2JS^2 -J'S^2 - JS \sum_{\vec{k}} \mathcal{C} (\vec{k}) \\
\label{magnetization-discrete}
M & = & \left< S_{i}^{1z} \right>_{\rm{o}} 
    = S - \left< c_{i}^{+} c_{i}^{} \right>_{\rm{o}} 
    = S - \frac{1}{N} \sum_{\vec{k}} v_{\alpha}^{2} (\vec{k})
\end{eqnarray}
\noindent
with
$\mathcal{C}_{k} = \frac{1}{2} (\epsilon_\alpha + \epsilon_\beta + 
(f-g) (u_{\alpha}^{2} - u_{\beta}^{2}) - 
(f+g))
$
and where the coefficients of the Bogoliubov transformation are
$u_{\alpha} = h/\sqrt{h^2 - (\epsilon_\alpha - f)^2}$, 
$v_{\alpha} = (f - \epsilon_\alpha)/\sqrt{h^2 - (\epsilon_\alpha + f)^2}$,
$u_{\beta} = (\epsilon_\beta + f)/\sqrt{(\epsilon_\beta + f)^2 - h^2}$ and
$v_{\beta} = h/\sqrt{(\epsilon_\beta + f)^2 - h^2}$.
\section{Dyson-Maleev Representation
\label{dysonmaleev}}
We use the same hamiltonian of Eq. \ref{hamiltonian} and introduce the Dyson-Maleev
transformation \cite{dyson56}.
Because of its algebra, the Dyson-Maleev transformation rewrites the hamiltonian of 
Eq. \ref{hamiltonian} as a sum of two and four operators terms.
The simplest method to solve the problem is to decouple the four operators term
introducing an {\it a priori} decoupling scheme.
In the present case, we follow the mean field decoupling of Ref. \onlinecite{chu91}.
Three quantities are introduced that correspond each to a two bosons like excitations,
\begin{eqnarray}
\mathcal{D} & = & \left<  a_{i}^{+}a_{i}^{}  \right> = \left< b_{j}^{+}b_{j}^{} \right>  \nonumber \\
\mathcal{O} & = & \left<  a_{i}^{+}a_{l}^{}  \right> = \left< b_{j}^{+}b_{m}^{} \right>  
\; \; ; \; \; i \neq l \; \; , \; \;j \neq m \\
\mathcal{T} & = & \left<  a_{i}^{+}b_{j}^{+} \right>   \nonumber
\end{eqnarray}
\noindent
Other types of excitations are neglected and considered as higher energy processes as
previously proposed by Takahashi \cite{takahashi89},
Using these approximations, the Dyson Maleev hamiltonian is then linearized 
and diagonalized through another Bogoliubov transformation.
The structure of the transformation is identical to the one of the linear spin
wave case, but with new coefficients.
These coefficients are 
$f_{k}^{'} = 2A + r B (\epsilon -1)$, 
$g_{k}^{'} = 2A + r B (\eta -1)$,
$h_{k}^{'} = A (\delta + \gamma)$ where 
$A = 1 - (\mathcal{D}+\mathcal{O})/2S$ and 
$B = 1 - (\mathcal{D}-\mathcal{T})/2S$.
Finally the three self consistent coefficients at zero temperature are obtained through the 
three coupled relations,
\begin{eqnarray}
\label{self-equations}
\mathcal{D} & = & \frac{1}{\rm N} \sum_k v_{\alpha}^{'2} (k) \nonumber \\
\mathcal{O} & = & - \frac{1}{\rm 2N} \sum_k u_{\alpha}^{'} (k) v_{\alpha}^{'} (k) (\delta + \gamma) \\ 
\mathcal{T} & = & \frac{1}{\rm 2N} \sum_k v_{\alpha}^{'2} (k) (\epsilon + \eta) \nonumber
\end{eqnarray}
\noindent
and the local magnetization of the N\'eel state is simply $M = \left< S_{i}^{1z} \right>_{0} = S - \mathcal{D}$.
\section{Results of the spin wave approach
\label{results-sw}}
\begin{figure}
\includegraphics[width=7.5cm]{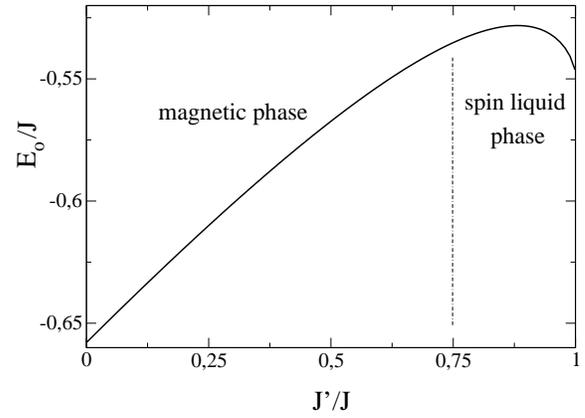}%
\caption{Energy per site as a function of the ratio $r=J'/J$, i.e
between the limits of the square lattice and the checkerboard lattice, for 
$S=1/2$. Beyond the critical ratio $r \approx 0.75$, the value of the
energy is meaningless as it describes the energy of an unstable state.
\vspace{0.5cm}
\label{energy-vs-r}}
\end{figure}

\begin{figure}
\includegraphics[width=7.5cm]{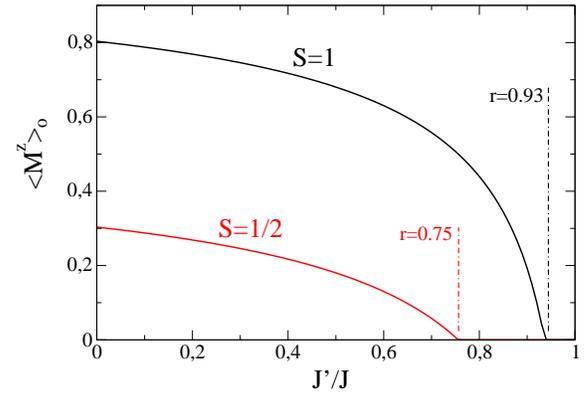}%
\caption{Local magnetization as a function of the ratio $r=J'/J$, i.e
from the square lattice to the checkerboard lattice. Red lines correspond
to the calculations made for $S=1/2$ and black lines for $S=1$.
\label{magnetization-vs-r}}
\end{figure}

\begin{figure}
\includegraphics[width=7cm]{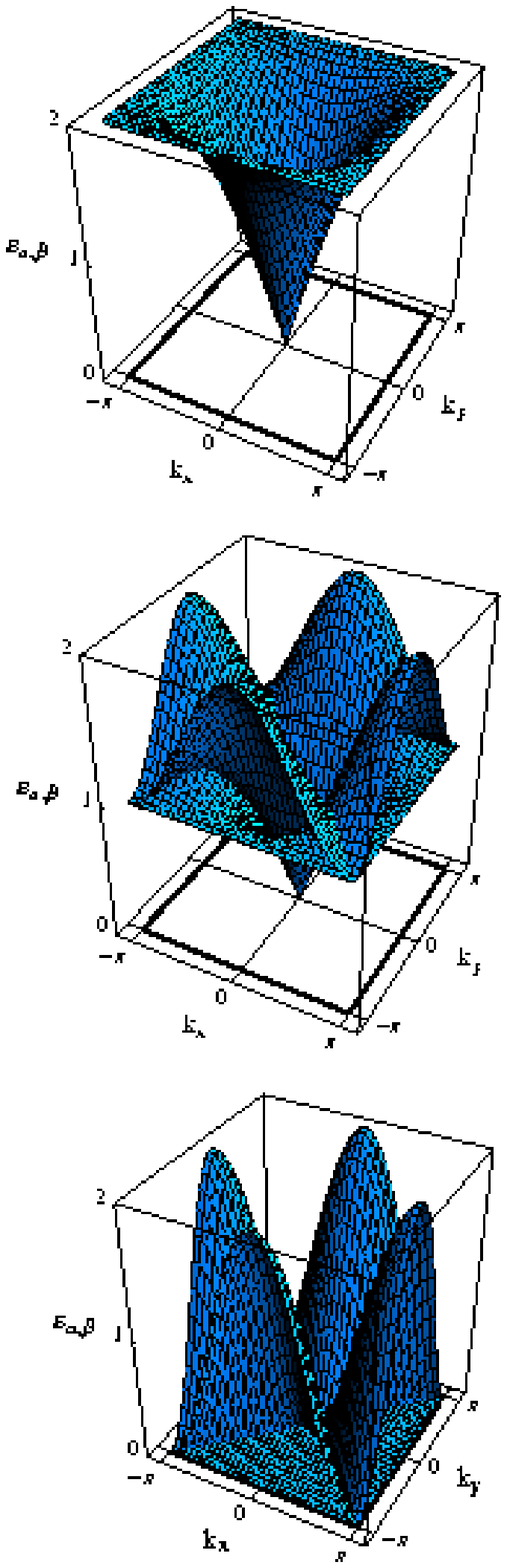}%
\vspace{+0.1cm}
\caption{Magnons dispersions $\epsilon_{\alpha,\beta} (k)$ 
over the Brillouin zone.
From top to bottom, $r=0,0.5,1$. At $r=1$, the lowest
branch is non-dispersive. This means that this point is
remarkable and reveals (as in the kagom\'e and the pyrochlore
lattice case) the presence of a soft mode related to local
degrees of freedom.
\label{magnons-dispersions}}
\end{figure}

\begin{figure}
\includegraphics[width=7.7cm]{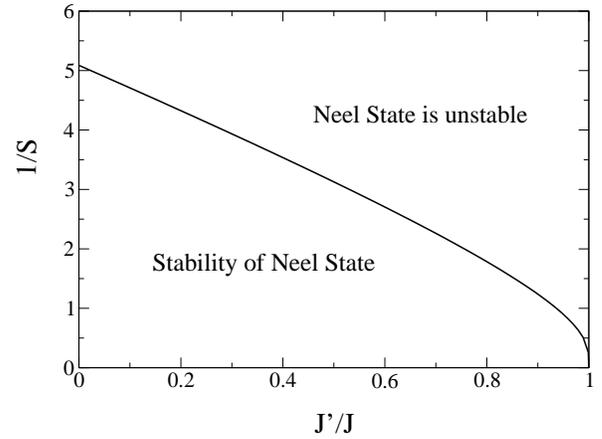}%
\caption{Stability of the Neel state. 
The straight line indicates the threshold below wich the Neel 
state is stable for a given ratio $r=J'/J$. Over this line, it is 
destabilized by geometry and quantum fluctuations. 
For $r=1$, it is always unstable, whatever the value of the spin $S$.
\label{phasediagram}}
\end{figure}

In the following, we fix $J=-1$ and take $-1 \le J' \le 0$.
We first discuss the linear spin wave results and then the Dyson-Maleev
approach.
In Fig.~\ref{energy-vs-r} and \ref{magnetization-vs-r}, the energy as well
as the local magnetization are plotted as functions of the ratio $r=J'/J$.
For the case $r=0$, well known results for the $S=1/2$ square lattice 
are recovered : $E=-0.658$ and $M=0.3$.
The magnon spectrum is very simple; there is only one dispersive branch,
which is two times degenerate in our case because of the 2-spin unit cell
description (Fig.~\ref{magnons-dispersions} top).
With increasing $r$, there are three zones to distinguish.%
First, in the range $0 \leq r \leq r_c$, the magnon spectrum has two
distinct manifolds, each of these being dispersive 
(Fig.~\ref{magnons-dispersions} middle).
This occurs now because the site of the unit cell no longer
have the same symmetry.
$r_c$ corresponds to a critical coupling below which the local
magnetization is finite, even renormalized by quantum corrections.
For $S=1/2$, $r_c \approx 0.75$ which was already obtained in
Ref. \onlinecite{singh98} and for $S=1$, $r_c \approx 0.93$.
Critical values for higher spins are shown in the phase diagram
of Fig.~\ref{phasediagram}.
The second zone corresponds to the range $r_c \leq r < 1$.
Because of quantum corrections, local magnetization is renormalized
to zero, which means that in this approximation, the system does
not show magnetic ordering.
Even if quantum fluctuations drive the system to disorder, one notes
that magnons are still dispersive.
This means that there is no peculiar contribution coming from their
structure.
This is no longer the case for the special point $r=1$.
At this point, the lowest branch of spin waves
is flat over the whole Brillouin zone 
(Fig.~\ref{magnons-dispersions} bottom) .
This has several consequences.
First, this means that at the level of this approximation, there will
always be a critical coupling, whatever the value of the spin.
This comes directly from the non-analytic behavior of $v_{\alpha} (\vec{k})$
when $r \rightarrow 1$ along directions $ky=\pm kx$.
Because of that the local magnetization diverges as $-1/\sqrt(1-r)$, 
i.e is renormalized to zero, whatever the value of S.
Consequently, we obtain a phase diagram ($1/S$,$J'/J$) very similar to the
one of the $J_1-J_2$ square lattice model where $J'/J$ plays the role
of $J_2/J_1$ (see Fig.~\ref{phasediagram} and Fig.~2 of Ref. \onlinecite{chandra88}).
This problem of flat mode is often encountered in frustrated lattices and
it rules out the possibility of studying the $r=1$ point without
going to next order of expansion in our model, i.e, without
including quartic terms.
Whether such an inclusion of quartic terms is relevant is not clear
as it corresponds to consider quartic modes as perturbations
of {\it zero} quadratic modes.
Nevertheless, it is the simplest next step to be done in the frame
of spin wave (Holstein-Primakoff) analysis \cite{chubukov92}.
Second, it has a meaning very close to MF conclusions.
If the lowest branch of magnons is flat, this means that the
system has no spin stiffness.
Thus, any local deformation will not propagate through lattice.
This is very close to the argument of continuous degeneracy
obtained through the MF approach, i.e the existence of 
an infinite number of local degrees of freedom 
(see Eq. \ref{constraints}).
These degrees of freedom are one of the signatures of 
frustrated systems and may be counted, in the case of classical
spins, through the ``Maxwellian counting'' rule \cite{moecha98}.
At the level of the LSW approximation, this means that quantum fluctuations
and geometrical frustration may destabilize the N\'eel state for a finite
range of $J'$.
Wether a quantum spin liquid ground state in this range is stable is
beyond this approach even if the existence of critical couplings goes
in the right direction.

\begin{figure}
\vspace{1cm}
\includegraphics[width=8cm]{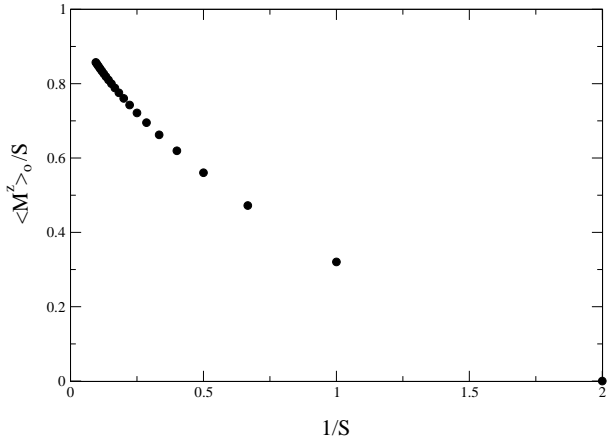}%
\caption{Relative local magnetization of the N\'eel state versus the 
inverse spin value $1/S$ at the isotropic point.
For all $S \ge 1$, the local magnetization is finite, suggesting that 
the ground state could be the N\'eel state.
For $S=1/2$, the local magnetization is renormalized to zero by quantum
fluctuations.
\label{localMzversus1/S}}
\end{figure}

\begin{figure}
\includegraphics[width=8cm]{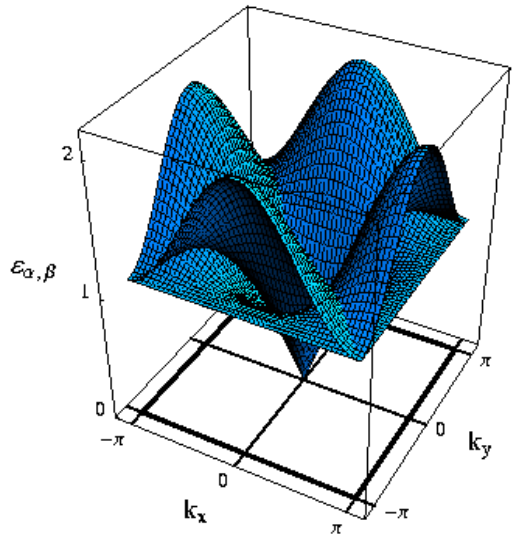}%
\caption{Energy dispersions corresponding to the bosons excitations around
the N\'eel state for the $S=1/2$ case in the limit $r \rightarrow 1$.
The dispersions are no longer pathological as they are in the LSW approximation
and the flat mode is destroyed.
This clearly indicates that destabilization of the N\'eel state is driven
here only by quantum fluctuations.
\label{dysondispersion}}
\end{figure}

We now turn to the Dyson-Maleev approach.
As for the preceding case, the local magnetization has been computed
for different value of the spin.
The main result is that for all $S \ge 1$, there is no critical coupling,
i.e the local magnetization of the N\'eel state is always finite in the
isotropic limit $r = 1$.
This is illustratred in Fig.~\ref{localMzversus1/S} where reduced local 
magnetization $\left< S_{i}^{z} \right>_{0} / S$ is plotted versus the inverse 
spin value $1/S$ at $r \rightarrow 1$.
The only exception is for $S = \frac12$ for which it seems that there is
a critical coupling at $r \approx 0.98$.
To test wether the stability of the N\'eel state for the $S = \frac12$
comes from a flat mode or from quantum fluctuations, we have plotted
the associated dispersions over the whole Brillouin zone 
(see Fig.~\ref{dysondispersion}). 
From the structure of the bosons excitations, it is clear that including quantum
fluctuations via the Dyson-Maleev formalism destroys the pathological
structure of the flat mode.
Therefore, destabilization of the N\'eel state is only driven by 
quantum fluctuations within this approximation and not by a zero spin
stiffness as it was the case in the LSW approximation in the limit 
$r \rightarrow 1$.
%

%
%
These results are quite surprising as they are qualitatively different
from the LSW approximation.
This suggests that the more accurately the quantum fluctuations are taken into account,
the more the N\'eel state is stable, even in the isotropic limit.
This is unexpected as the checkerboard antiferromagnet was expected to behave like
the kagom\'e antiferromagnet.
In the latter case, the reduction of the local magnetization to zero by 
quantum fluctuations and geometry, within a finite range of couplings, 
has been shown using two distinct numerical methods \cite{wallech97}.
It was found that going from the triangular limit to the kagom\'e
limit, there is a singular point associated to a critical 
coupling ($r_c = J'_{c}/J \approx 0.9$), where local magnetization
vanishes.
%
%
Here, we obtain the same conclusion within the LSW approximation but not
within the mean field treatment of the Dyson-Maleev approach.
Wether this discrepency comes from the quantum fluctuation that are better
incorporated in the latter formalism, or comes from the mean field treatment
of the DM representation is not clear.
Nevertheless, it is interesting to note that a recent work \cite{moessner2001}
shows the same tendency, i.e a transition from an $S=0$ ground state to a stable
N\'eel state with increasing spin size $S$.
If the present calculations treats correctly quantum fluctuations (which is not
proven in this paper), the frontier between the expected VBS state and the
N\'eel ordering \cite{moessner2001} would then be at $S=1$.
Nevertheless, it should be stressed that the present work does not test
wether other classical ground states could be stable at the isotropic 
point and have the same or a different energy then the N\'eel ground state.
Consequently, an intrinsic or accidental degeneracy of magnetic ground
states for $S \ge 1$ is not ruled out by this approach.
\section{The $S=1/2$ case
\label{s-small}}
\begin{figure}
\includegraphics[width=8cm]{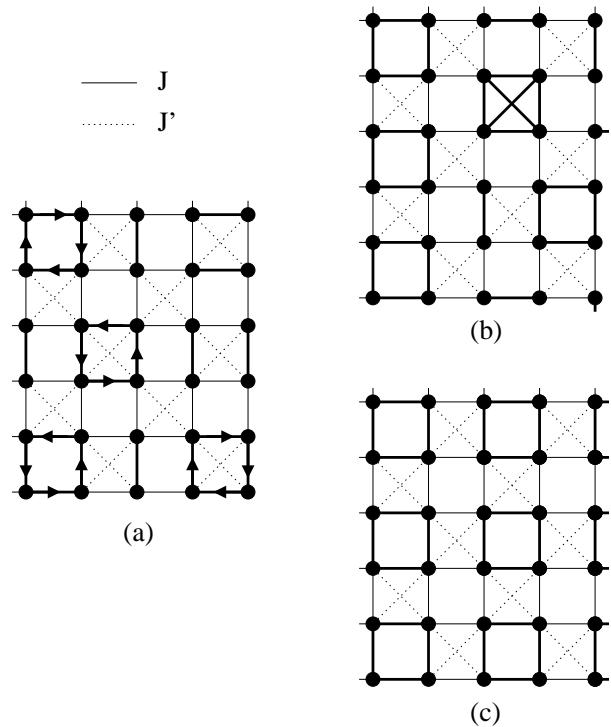}%
\caption{(a) General box phase on the $J$ square lattice of
the anisotropic checkerboard antiferromagnet 
(as shown in Fig.~4 of Ref. \onlinecite{rokhsar90}). 
Bold lines without arrows correspond
to non zero $\chi_{ij}$ while thin lines, plain or dashed, correspond to
$\chi_{ij} = 0$. 
Bold lines with arrows indicate that $\chi_{ij} = \chi e^{i \pi / 4}$
where $\chi$ is real.
The product of $\chi_{ij}$ around each box is negative, indicating a flux of
$\pi$.
These boxes are the large-n equivalent of a 4-spins singlet on a square.
(b) Picture of a general box phase for the $S=1/2$ case.
Bold squares correspond to 4-spins singlets and bold lines to 2-spins singlets.
If $J'$ interactions are within a box, they must be taken into account and
are also in bold as they enter the calculation of the energy.
The different energies for squares with crossings, squares without crossings 
and links are respectively $E = -1.5$, $E = -2$ and $E = -0.75$.
(c) Picture of the valence bond solid for the $S=1/2$ case, expected to
be the ground state.
This state is two times degenerate and involves only squares without 
crossings.
\label{large-n-solutions}}
\end{figure}

As shown in the preceding sections, it seems that only for $S=1/2$ can the
ground state be non-magnetic.
To shed light on the behavior of the $S=1/2$ checkerboard 
antiferromagnet, it is possible to use a ``fermionic'' SU(n) generalization
of the SU(2) group
[Note that $n>2$ does not correspond to a higher spin representation of SU(2).
Therefore, a mean-field solution of the SU(n) hamiltonian is expected to be
much closer to the SU(2) physics than the classical description of 
sec. \ref{sw-hamiltonian}].
This formalism has been previously introduced by Marston and Affleck 
\cite{marston89} and applied to the square antiferromagnet.
The advantage of this description is that it works on all lattices,
even frustrated \cite{marston91} or not bipartite, and that it is 
well suited for the checkerboard antiferromagnet as spin order never occurs in the exactly
solvable large-n limit \cite{marston89}, as it is expected in the 
range $r_c \le r \le 1$ for $n = 2$.
At large n, the hamiltonian can be written as
\begin{equation}
H = 	\sum_{(i,j)} \left\{ 
		\left(  \frac{n}{J} \right) \left| \chi_{ij}  \right|^2 + 
		\sum_{\alpha = 1}^{n} \left(  \chi_{ij} c_{i \alpha}^{+} c_{j \alpha}^{}   + \textrm{H.c} \right)	
		     \right\}
\end{equation}
where $\chi_{ij} = (J/n) c_{i \alpha}^{+} c_{j \alpha}^{}$ are bond variables
and where spin-spin interaction has been expressed by use of electron operators through
$S_i . S_j = (1/2) \sum_{\alpha,\beta} c_{i \alpha}^{+} c_{j \beta}^{} c_{j \beta}^{+} c_{j \alpha}^{} + \textrm{constant}$.
The $\chi$ variables may be seen as valence bond operators \cite{affleck88} as 
$(n/J) \chi_{ij}^{+} \chi_{ij}$ is the number of valence bonds on the
link $(ij)$.
Therefore, in this formalism, every ground state is described by a map
of these complex scalar fields.
Several authors have concentrated on the mean-field solutions (``$1/n = 0$'')
\cite{marston89,dombre89,rokhsar90} and in particular, it is possible to apply
here a theorem established by D. Rokhsar \cite{rokhsar90}.
First it is clear that the checkerboard antiferromagnet is ``dimerizeable with respect to $J$'' as 
``it is possible to partition the network into disjoint pairs of sites such
that (a) every site belongs to one and only one pair, and (b) for each pair
(i,j) whe have $J_{ij} = J$''. 
In the range $J \ge J'$ it is possible to apply his theorem which shows
that mean-field ground states are (spin-Peierls or box)-phases \cite{rokhsar90}
(see Fig.~\ref{large-n-solutions}).
Keeping in mind that the case of interest is for $n = 2$, wee look for the
proposed states that lower the energy of the real $S=1/2$ case.
It is straightforward to see that box-phases lower this energy with 
respect to $J$, as soon as these phases are regular tilings of the 
underlying square lattice by disconnected $J$-squares 
(see Fig.~\ref{large-n-solutions}).
Taking now into account the $J'$ couplings, only one of these phases
survives, where $J'$ links are not frustrated.
At this point, it means that {\it if} the SU(n) description of the SU(2)
 case is relevant, it suggests that for $S=1/2$, in the range 
$r_c \le r \le 1$, the ground state should be two times degenerate (because
there are two choices for the tiling), consisting in a valence bond solid 
of disconnected squares without crossings (in the range $0 \le r < r_c$ it is clear that 
this formalism is not relevant as the local magnetisation is finite).
The proposed ground state (see Fig.~\ref{large-n-solutions}(c)) could be tested by comparing the VBS energy
per site $E = -0.5$ and the correlations functions with results of exact 
diagonalisations for example \cite{fouet01}.
\section{Discussion and Conclusion
\label{conclusion}}
To conclude, Linear Spin Wave analysis has shown that away from
the isotropic checkerboard lattice limit, there is a finite
range of coupling for wich the N\'eel state is unstable, whatever
the value of the spin.
Strikingly, when using a self consistent decoupling within the Dyson-Maleev
formalism, it is shown that the N\'eel state is never destabilized,
excepted for the $S = \frac12$ case.
The stability of the N\'eel state with respect to other magnetized
states has not been tested in this work.
Recent work \cite{moessner2001} supports that this state should be
the ground state for large $S$.
It is compatible with the present results although it is at variance
with results obtained with the infinite-component \cite{canals01bis}
antiferromagnetic model on the chekerboard lattice ($d \rightarrow \infty$).
A possible explanation for this \cite{discuss} is that for large S, there should be quantum 
order by disorder (as supported in \cite{moessner2001}) but this
order by disorder comes with an energy scale of J/S rather than J.
Therefore, in the classical limit, the temperature at which the
correlation length grows exponentially goes to zero.
In other words, the limits $T \rightarrow 0$ and $S \rightarrow \infty$ 
(or $d \rightarrow \infty$) should not commute.
For the special $S = \frac12$ case, the fermionic large-n limit suggests
that the ground state is a valence bond solid of disconnected 4-spins
singlets on squares without crossings.
That conclusion is consistent with results of Refs. \onlinecite{moessner2001}
and \onlinecite{fouet01}.

The present work shows that the isotropic checkerboard antiferromagnet
should display behaviors qualitatively different from the one of
the kagom\'e and the pyrochlore antiferromagnet.
Moreover, quantum fluctuations have to be treated with care as
they can drive the system from a purely quantum ground state to
a N\'eel like ground state.
In this frame, the structure of the singlet and triplet towers
of state should be extremally $S$-dependent and exact diagonalization
of finite clusters with different spin size would be of high 
interest, especially to test wether the descrepency already
appears between the $S = \frac12$ and the $S = 1$ cases.
%
%
%
 
%
%
%
\begin{acknowledgments}
The author would like to thank Maged Elhajal, Peter Holdsworth, Claudine Lacroix and Roderich Moessner 
for interesting and stimulating discussions.
\end{acknowledgments}


\begin{thebibliography}{10}
%

\bibitem{haldane}
F. D. M. Haldane,
Phys. Lett. A {\bf 93}, 464 (1983).

\bibitem{wannier50}
G. H. Wannier,
Phys. Rev. {\bf 79}, 357 (1950).

\bibitem{houtappel50}
R. M. F. Houtappel,
Physica {\bf 16}, 425 (1950).

\bibitem{liebmattis62}
E. H. Lieb and D. Mattis,
J. Math. Phys. (N.Y.) {\bf 3}, 749 (1962).

\bibitem{leche97}
P. Lecheminant, B. Bernu, C. Lhuillier, L. Pierre and P. Sindzingre,
Phys. Rev. B {\bf 56}, 2521 (1997).

\bibitem{garcan99}
D. A. Garanin and B. Canals, 
Phys. Rev. B {\bf 59},  443  (1999).

\bibitem{moecha98} 
R. Moessner and J. T. Chalker, 
Phys. Rev. Lett. {\bf 80}, 2929 (1998);
R. Moessner and J. T. Chalker, 
Phys. Rev. B {\bf 58}, 12049 (1998).
R. Moessner,
{\it D. Phil. thesis} 
(Oxford University, 1996)

\bibitem{canals9800}
B. Canals and C. Lacroix, 
Phys. Rev. Lett. {\bf 80},  2933  (1998);
B. Canals and C. Lacroix, 
Phys. Rev. B {\bf 61},  1149  (2000).


\bibitem{diep94}
For a collection of articles, see
{\it Magnetic systems with Competing Interactions : Frustated Spin System}, 
edited by H. T. Diep (World Scientific, Singapore, 1994); for Ising systems, 
see R. Liebmann, 
{\it Statistical Mechanics of Periodic Frustrated Ising Systems} 
(Springer, Berlin, 1986); 
for reviews, see 
A. P. Ramirez, 
Annu. Rev. Mater. Sci. {\bf 24}, 453, (1994); 
P. Schiffer and A. P. Ramirez, 
Comments Cond. Mat. Phys. {\bf 18}, 21, (1996);
M. J. Harris and M. P. Zinkin, 
Mod. Phys. Lett. B {\bf 10}, 417, (1996).


\bibitem{lieb99}
E. H. Lieb and P. Schupp,
Phys. Rev. Lett. {\bf 83}, 5362 (1999).

\bibitem{chandra88}
P. Chandra and B. Dou\c cot,
Phys. Rev. B {\bf 38}, 9335 (1988).

\bibitem{dyson56}
F. J. Dyson, Phys. Rev. {\bf 102}, 1217 (1956); 
F. J. Dyson, Phys. Rev. {\bf 102}, 1230 (1956);
S. V. Maleev, Zh. Eksp. Teor. Fiz. {\bf 30}, 1010 (1957) 
[Sov. Phys. JETP {\bf 6}, 776 (1958)].

\bibitem{rei91}
E.F. Bertaut, 
in {\it Spin Arrangement and Crystal Structure, Domains and Micromagnets}, 
edited by T. Rado and H. Suhl, 
Magnetism Vol. III (Academic Press, New York, 1963);
J. N. Reimers, A. J. Berlinsky and A.-C. Shi., 
Phys. Rev. B
{\bf 43} (1991) 865.

\bibitem{chubukov92}
A. Chubukov, 
Phys. Rev. Lett. {\bf 69}, 832 (1992).

\bibitem{chu91}
D. Chu and J. L. Shen, Phys. Rev. B {\bf 44}, 4689 (1991).

\bibitem{takahashi89}
M. Takahashi, Phys. Rev. B {\bf 40}, 2494 (1989).

\bibitem{singh98}
R.R.P.Singh, O.A.Starykh and P.J.Freitas, 
J. Appl. Phys. {\bf 83}, 7387 (1998).

\bibitem{wallech97}
D. J. J. Farnell, R. F. Bishop and K. A. Gernoth,
cond-mat/0010477.
C. Waldtmann, P. Lecheminant, {\it unpublished}.

\bibitem{moessner2001}
R. Moessner, Oleg Tchernyshyov and S. L. Sondhi, cond-mat/0106288.

\bibitem{marston89}
J. B. Marston and I. Affleck, 
Phys. Rev. B {\bf 39}, 11538 (1989).

\bibitem{marston91}
J. B. Marston and C. Zeng, 
J. Appl. Phys. {\bf 69}, 5962 (1991).

\bibitem{affleck88}
Ian Affleck and J. Brad Marston, 
Phys. Rev. B {\bf 37}, 3774 (1988).

\bibitem{dombre89}
Thierry Dombre and Gabriel Kotliar,
Phys. Rev. B {\bf 39}, 855 (1989).

\bibitem{rokhsar90}
Daniel S. Rokhsar,
Phys. Rev. B {\bf 42}, 2526 (1990).

\bibitem{fouet01}
J.-B. Fouet, M. Mambrini, P. Sindzingre and C. Lhuillier,
condmat/0108070.

\bibitem{canals01bis}
B. Canals and D. A. Garanin,
cond-mat/0102235.

\bibitem{discuss}
R. Moessner, private communication.



\end{thebibliography}
\end{document}